\documentclass[sn-mathphys,Numbered]{sn-jnl}% Math and Physical Sciences Reference Style
\usepackage{graphicx}%
\usepackage{multirow}%
\usepackage{amsmath,amssymb,amsfonts}%
\usepackage{amsthm}%
\usepackage{mathrsfs}%
\usepackage[title]{appendix}%
\usepackage{xcolor}%
\usepackage{textcomp}%
\usepackage{manyfoot}%
\usepackage{booktabs}%
\usepackage{algorithm}%
\usepackage{algorithmicx}%
\usepackage{algpseudocode}%
\usepackage{listings}%
\usepackage{multibib}
\usepackage{caption}

\theoremstyle{thmstyleone}%
\theoremstyle{thmstyletwo}%

\theoremstyle{thmstylethree}%

\begin{document}

\title[Article Title]{Estimating and approaching maximum information rate of noninvasive visual brain-computer interface}


\author[1]{\fnm{Nanlin} \sur{Shi}}\email{shinl.diego@gmail.com}

\author[1]{\fnm{Yining} \sur{Miao}}\email{myn23@mails.tsinghua.edu.cn}

\author[1]{\fnm{Changxing} \sur{Huang}}\email{huangcx23@mails.tsinghua.edu.cn}

\author[1]{\fnm{Xiang} \sur{Li}}\email{l-xiang16@mails.tsinghua.edu.cn}

\author[1]{\fnm{Yonghao} \sur{Song}}\email{songyh22@mails.tsinghua.edu.cn}

\author[2]{\fnm{Xiaogang} \sur{Chen}}\email{chenxg@bme.cams.cn}

\author[3]{\fnm{Yijun} \sur{Wang}}\email{wangyj@semi.ac.cn}

\author[*1]{\fnm{Xiaorong} \sur{Gao}}\email{gxr-dea@mail.tsinghua.edu.cn}

\affil*[1]{\orgdiv{Department of Biomedical Engineering}, \orgname{School of Medicine}, \orgaddress{\street{Tsinghua University}, \city{Beijing}, \postcode{100084},  \country{China}}}

\affil[2]{\orgdiv{Institute of Biomedical Engineering}, \orgname{Chinese
Academy of Medical Sciences and Peking Union Medical College}, \orgaddress{\street{Street}, \city{Tianjin}, \postcode{300192},  \country{China}}}

\affil[3]{\orgdiv{State Key Laboratory on Integrated Optoelectronics}, \orgname{Institute of Semiconductors},\orgname{Chinese Academy of Sciences} \city{Beijing}, \postcode{100083}, \country{China}}


\abstract{The mission of visual brain-computer interfaces (BCIs) is to enhance information transfer rate (ITR) to reach high speed towards real-life communication. Despite notable progress, noninvasive visual BCIs have encountered a plateau in ITRs, leaving it uncertain whether higher ITRs are achievable. In this study, we investigate the information rate limits of the primary visual channel to explore whether we can and how we should build visual BCI with higher information rate. Using information theory, we estimate a maximum achievable ITR of approximately 63 bits per second (bps) with a uniformly-distributed White Noise (WN) stimulus. Based on this discovery, we propose a broadband WN BCI approach that expands the utilization of stimulus bandwidth, in contrast to the current state-of-the-art visual BCI methods based on Steady-State Visual Evoked Potentials (SSVEPs). Through experimental validation, our broadband BCI outperforms the SSVEP BCI by an impressive margin of 7 bps, setting a new record of 50 bps. This achievement demonstrates the possibility of decoding 40 classes of noninvasive neural responses within a short duration of only 0.1 seconds. The information-theoretical framework introduced in this study provides valuable insights applicable to all sensory-evoked BCIs, making a significant step towards the development of next-generation human-machine interaction systems.}


\maketitle

\section{Introduction}\label{sec1}

Brain-computer interfaces (BCIs) strive to establish high information rate communication between the human brain and the external world\cite{anumanchipalli2019speech,willett2021high}. In the context of visual BCIs, this is accomplished by leveraging the primary visual pathway to transfer information through stimulus-evoked neural responses in non-invasive settings\cite{gao2014visual,chen2015high}. Over the past two decades, electroencephalogram (EEG)-based visual BCIs have enhanced the information transfer rate (ITR) through innovative paradigm designs\cite{chen2015high} and decoding algorithms\cite{nakanishi2017enhancing}, reaching a peak theoretical ITR of 16 bps. Despite these achievements, the full potential of utilizing the primary visual pathway remains uncertain. Consequently, an important question arises: is it possible to achieve further substantial improvements in the information rate, and if so, how can it be accomplished?

The information rate of visual BCI is constrained by the information processing capability of visual system. Extensive research has been conducted to investigate the information processing capability, particularly at the single neuron level, in the Retinal Ganglion Cells (RGC)\cite{warland1997decoding,passaglia2004information} and Lateral Geniculate Nucleus (LGN)\cite{yu2010estimating}. These studies typically employ dynamic broadband flicker stimuli to stimulate visual neurons and examine the maximum achievable information transfer within Shannon's theoretical information framework\cite{borst1999information,shannon1949communication}. Additionally, linear system modeling is employed to establish stimulus-response functions that describe the visual encoding process as a linear convolution process\cite{benardete1997receptive,deangelis1995receptive}. These stimulus-response functions, often represented by the temporal receptive field (TRF) or temporal response function, provide insights into the relationship between the stimulus and response\cite{crosse2016multivariate}. By combining the analysis of maximum information transfer with the stimulus-response function, information theory allows us to quantify the mutual information between the stimulus and response\cite{borst1999information}.

Although the maximum information transfer of the primary visual pathway remains uncertain, visual BCIs have still achieved high information rates through innovative paradigm designs and decoding algorithms, particularly in the field of Steady-State Visual Evoked Potential (SSVEP)-based BCIs where each target is modulated by a narrowband stimulation on spectrum. In 2015, Chen et al. introduced the SSVEP paradigm called Joint Frequency Phase Modulation (JFPM), which encoded 40 targets within the frequency range of 8-15.8 Hz\cite{chen2015high}. Each target's contrast was modulated by different frequencies and phases of a sine wave. Building upon the JFPM paradigm, Nakanishi et al. developed the Task-related Component Analysis (TRCA) algorithm, which utilized individual neural responses as template matching knowledge\cite{nakanishi2017enhancing}. Combined with the JFPM paradigm, this algorithm achieved the highest ITR of 376 bits per minute (bpm), equivalent to 16 bits per second (bps) of mutual information measure by theoretical ITR (ITR*). However, since then, the information transfer rate has not experienced significant growth.

Has the current best narrowband SSVEP BCI reached the information transfer limits of the primary visual pathway? If not, how can we further enhance mutual information transfer to enable more efficient interaction with the human brain? To address these questions, we first introduced the information theory framework and modeled the primary visual channel as a communication channel. Based on the framework and related studies, we estimated the average channel capacity to be 63 bit per second (bps) using the broadband stimulus modulated by spatially uniformed White Noise (WN)\cite{chichilnisky2001simple}. Consequently, we observe that the JFPM SSVEP paradigm, which stimulates frequencies between 8-15.8 Hz, does not fully utilize the spectrum resources of the visual channel, and thus does not approach the limits of mutual information. To enhance mutual information, we propose a novel broadband BCI based on WN stimulation ranging from 1-30 Hz. The broadband WN BCI achieves a peak theoretical ITR of 50 bps, surpassing the previous record of 16 bps. The effectiveness of the novel broadband visual BCI paradigm is validated through offline and online BCI experiments. Additionally, we analyze six more relevant datasets from the perspectives of information source, channel and receiver, aiming to explore further opportunities for developing high-speed BCIs. Overall, this study introduced the information theory framework to estimates the theoretical limits of noninvasive visual BCIs. Further, the framework guides the exploration of spectrum resources by expanding the stimulation bandwidth using a broadband WN paradigm, resulting in a new record of mutual information. The information theory framework can be applied to all sensory evoked BCIs, potentially paving the way for a practical and reliable communication pathway for both healthy individuals and those with disabilities.

\section{Results}\label{sec2}

\subsection{Estimating information rate bounds of primary visual pathway}

\begin{figure}[!htbp]
    \centering
    \includegraphics[width=\textwidth]{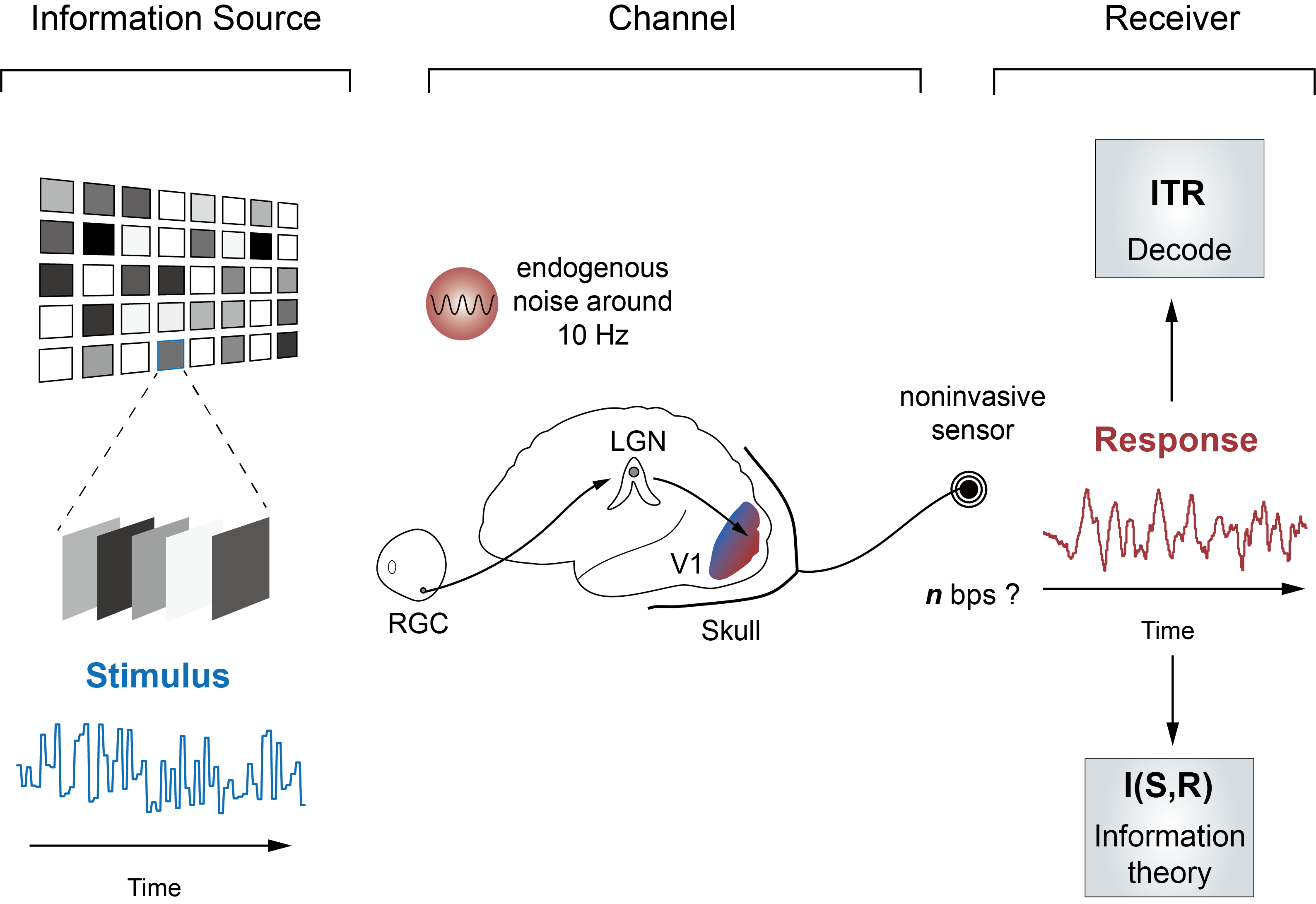}
    \caption{The primary visual pathway as information channel. The temporal dynamic stimulus is treated as source, the primary visual pathway consists of RGC-LGN-primary visual cortex is the channel, and the noninvasive sensor is the information receiver.}
    \label{fig:enter-label}
\end{figure}
In a typical visual BCI system, participants fixate their gaze on a specific stimulus that is modulated by a unique sequence. The mutual information contained in the stimulus ($S$) and response ($R$) can be quantified either through decoded ITR or using information theory as I(S,R), both measured in bps (see Fig. 1). To estimate the information transfer in this process, we define the dynamic stimulus sequence as information source, the primary visual pathway as the information channel, and the noninvasive sensors as the information receiver. 

Consistent with prior research\cite{passaglia2004information,chichilnisky2001simple,buracas1998efficient,van2001information}, our proposed model for the information channel adopts the additive Gaussian channel hypothesis (Fig. 2b). This hypothesis assumes that the signal is subject to interference from independent colored noise. Under this additive assumption, the mutual information is determined by the Signal-to-Noise Ratio SNR(f) in the frequency domain\cite{shannon1949communication}. The symmetric nature of mutual information permits two perspectives for calculating the SNR(f). These encompass the upper bound method calculated from the response domain and the lower bound method calculated from the stimulus domain (see the information theoretical framework part in the Method)

To estimate these bounds, we recorded the EEG responses evoked by 160 classes of broadband spatially unified white noise stimuli ranging from 1-30 Hz (\textbf{\textit{n}}=10, see Method). The lower bound represents the information that can be captured by stimulus-response functions. Therefore, it calculates SNR(f) from the stimulus perspective, where the signal component is considered to be the model-reconstructed stimulus (depicted by the blue lines in Fig. 2a), and the disparity with the actual stimulus is treated as the additional noise. The spectral representation in Fig. 2d demonstrates that, from the stimulus perspective, the primary visual system acts as a channel that band-pass filters the broadband information, centering it around 10 Hz (blue line). The upper bound, also referred to as channel capacity, is calculated from the response domain. It involves considering the trial-averaged response as the signal component and the residual as the additive noise, as illustrated in Fig. 2c. As depicted in the spectral representation in Fig. 2e, we can still observe the band-pass effect centered around 10 Hz. However, the background noise in the upper bound significantly differs from that in the lower bound. The additive noise of the upper bound also exhibits oscillatory behavior at around 10 Hz and can be observed in each stimulus class.
\begin{figure}[!htbp]
    \centering
    \includegraphics[width=\textwidth]{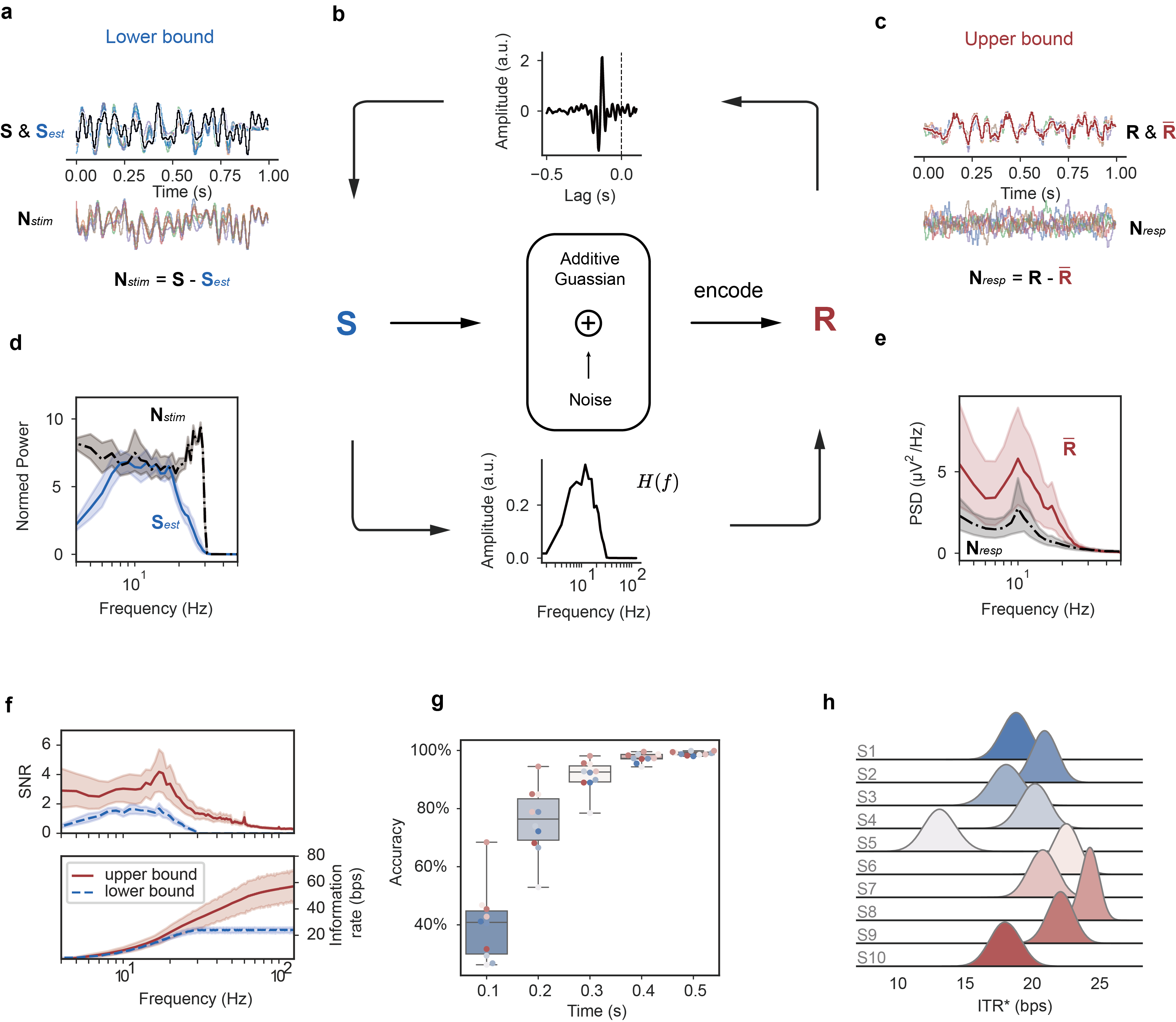}
    \caption{Estimate information rate of from channel model. \textbf{a-b}, The temporal and spectral representation of lower bound signal and noise component. \textbf{c}, The schematic diagram of additive gaussian channel. The upper panel denotes the TRF of a representative participant, the lower panel means the frequency response $H(f)$ of the same participant. \textbf{d-e}, The temporal and spectral representation of upper bound signal and noise component. \textbf{f}, The subject-averaged SNR(f) and the resulting mutual information of upper and lower bound method (\textbf{\textit{n}}=10, mean, 95\% CI) \textbf{g}, The 160-class classification accuracy with the increase of time (\textbf{\textit{n}}=10, medians, interquartile ranges, maxima and minima) \textbf{h}, The distribution of theoretical ITR* at 0.2 s for 10 participant. }
    \label{fig:enter-label}
\end{figure}
The distinction of noise and signal distribution leads to distinct SNR(f) values. While the lower bound filters the information from 1-30 Hz, primarily centering around 10 Hz, the upper bound estimation peaks in the beta band due to the non-phase-locked endogenous noise. Reflected on mutual information, we estimated the information rate of 63±20 bps for the upper bound, and 25±3 bps for the lower, as in Fig. 2f.    The results are close to multiple studies conducted on a part of visual system of various animals (See Table S1 in Supplementary material). Notably, the two estimation methods yield similar results within the 1-20 Hz frequency range, indicating that linear modeling can effectively explain the information transfer primarily occurring in low-frequency bands. However, for frequencies above 30 Hz, the upper bound estimation continues to increase while the lower bound estimation plateaus.

With an average channel capacity of 63±20 bps, it is evident that the current reported performance record of SSVEP BCI falls significantly below half of this capacity. This observation suggests that the key to improve lies in the spectral resources represented by SNR(f). The upper bound SNR(f) indicates that under broadband stimulation, the information contribution across the entire spectral domain exceeds the range covered by the JFPM paradigm, which is limited to frequencies between 8 to 15.8 Hz. Consequently, the decoded BCI information rate has the potential to be significantly enhanced by expanding the stimulation frequency range.

Most importantly, we found that by individual-calibrated algorithm like TDCA(see decoding algorithm in Method), the between-class responses under WN stimulation can achieve accurate single-trial classification. The classification accuracy can reach up to an average of 91\% at 0.3 s for 160 classes (Fig. 2g), which translates to ITR* of 15.76 bps  . The finding confirms that broadband white noise can build a novel visual BCI paradigm, and potentially exceed the current record of narrowband SSVEP BCI. However, owing to the stochastic nature of broadband modulation, the arrangement of stimulus sequences necessitates optimization. For comparison with the state-of-the-art 40-target SSVEP speller, the optimization of 40 sequences from the entire coding space becomes essential. When selecting a batch of non-overlapping classes (\textbf{\textit{n}}=40) from the total of 160 classes, the resulting ITR* can exhibit notable variation (refer to Fig. 2h).  To test whether the proposed broadband paradigm can surpass the current best JFPM paradigm and possibly reached a new height in visual BCI record, we chose a two-step optimization method to find the customized best stimulus combination for each participant (see optimization part in Method). 

\subsection{Narrow and broadband paradigm BCIs}

\begin{figure}[!htbp]
    \centering
    \includegraphics[width=\textwidth]{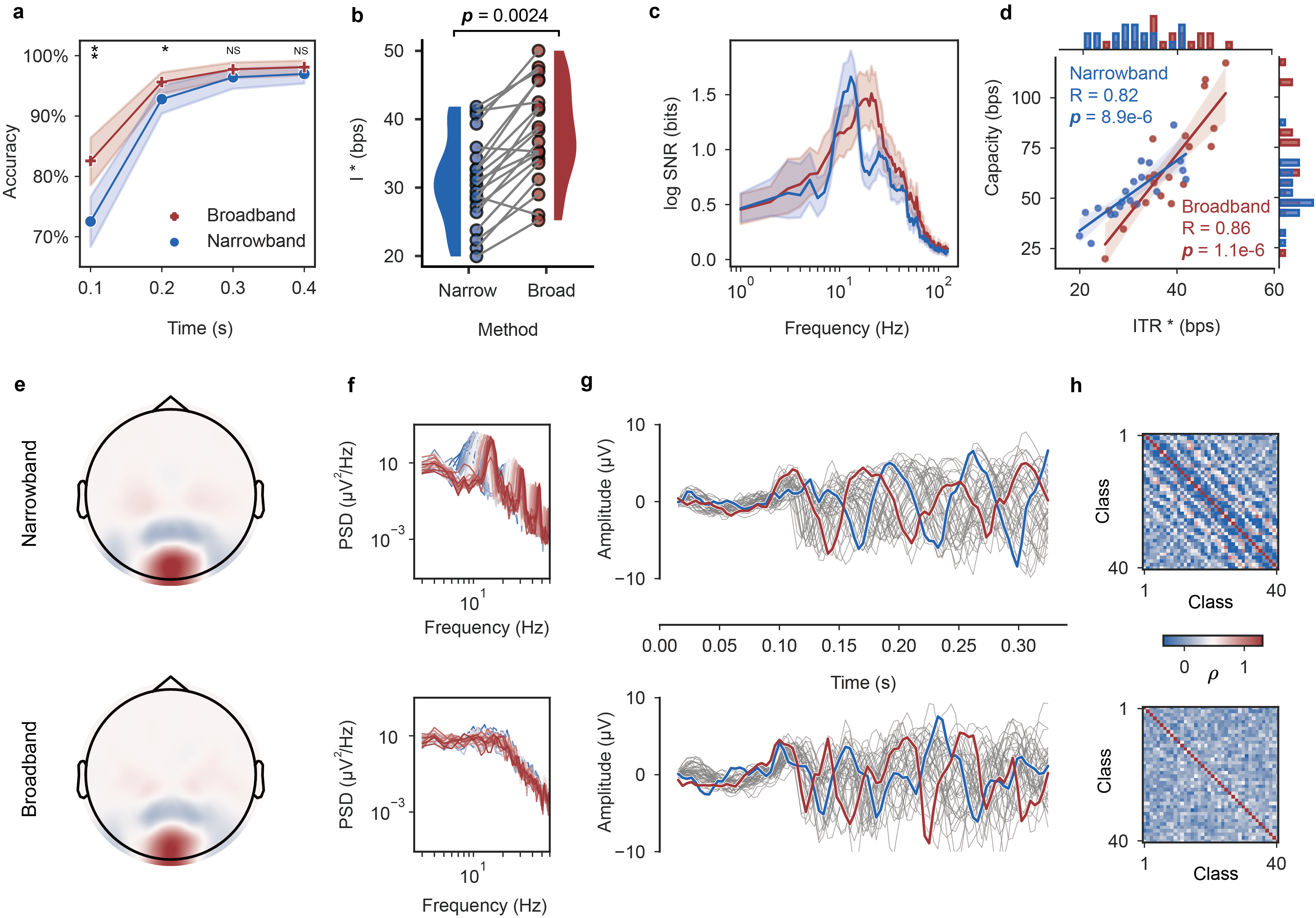}
    \caption{Performance and characteristics of broadband/narrowband visual BCI. \textbf{a}, The classification accuracy of two paradigms (\textit{n}=20, mean, 95\% CI as error bar, statistical significant (p=0.003 at 0.1 s, \textit{P}=0.01 at 0.2 s, NS when above 0.3 s, two-sided paired t-Test) \textbf{b}, The highest ITR* of two paradigms (\textit{n}=20, \textit{P}=0.0024, two-sided paired t-Test)   \textbf{c}, The log SNR(f) of two paradigms (mean, 95\% CI) \textbf{d}, The linear regression between ITR* and upper bound capacity (\textit{n}=20, error bar is 95\% CI, narrowband: R=0.82,p=8.9e-6, broadband: R=0.86, \textit{P}=1.1e-6 ) \textbf{e}, The spatial pattern of two paradigm, representative subject at 0.2 s. \textbf{f}, The spectral representation of group averaged evoked responses, each line from blue to red represents a stimulation class. \textbf{g}, The temporal evoked response of a representative subject, each line represent a class, red and blue lines are highlighted for demonstration. \textbf{h}, The Pearson-correlation coefficients between classes, group averaged at 0.1 s (\textit{n}=20). }
    \label{fig:enter-label}
\end{figure}
To validate whether the decoded information rate can be enhanced by broadband WN BCI, we conducted comparative experiments involving 20 healthy subjects. The results depicted in Fig. 3a demonstrate a notable improvement in decoding accuracy at 0.1 s, rising from 73\% for narrowband to 83\% for broadband (p=0.002, paired t-test). The highest decoded theoretical ITR also exhibits a significant increase from 31 bps to 38 bps (p=0.0024, paired t-test) for narrowband and broadband, respectively. This superiority of the broadband paradigm is further validated across various parameter spaces, including montages, training block sizes, and sub-bands (refer to Figures S4 and S5 in the Supplementary Material).The highest-performing participant achieved an ITR* of 50 bps , corresponding to a remarkable accuracy of 97\% at a time frame of only 0.1 s for the broadband paradigm (Fig. 3b). As time decreases, the ITR* for both paradigms continues to rise (refer to Figure S8 in the Supplementary Material). Nevertheless, the broadband paradigm's advantage becomes more pronounced. We then compare the theoretical ITR* with that of various related studies in Figure S8a, where the theoretical value reported in this study exhibits significant superiority over the values reported in the related studies . It is worth noting that the JFPM achieved an ITR* of 30 bps, surpassing the previous record of 16 bps obtained using the same paradigm\cite{chen2015high,nakanishi2017enhancing}. In comparison to earlier studies, the enhancements observed in JFPM within this study can be attributed to several factors: the utilization of a greater number of electrodes that cover the entire partial-occipital lobe, the implementation of more advanced algorithms, and the inclusion of diverse demographic groups of JFPM in this study.

The improved decoding performance of broadband over narrowband BCI can be attributed the expanded coverage on the SNR(f) between two paradigms, shown in Fig. 3c, where the narrowband SNR(f) mainly distributed in the fundamental frequency of 8-15.8 Hz. The SNR(f) of these two also demonstrates the fundamental difference between narrow and broadband modulation. The narrowband stimulation elicits resonance effect at around 10 Hz, which means alpha band is the optimal frequency band to devise a narrowband BCI. However, the broadband stimulation is interfered with by 10 Hz endogenous noise, resulting in the SNR(f) peak at the beta band. This dual effect of phase-locked and non-phase-locked alpha power play an import role in transmitting information in primary visual channel. To investigate the relationship between channel capacity and the highest decoded ITR*, we examined the linear regression between these two mutual information metrics. The results in Fig. 3d revealed a strong correlated relationship between the two metrics (narrowband: R=0.82, \textit{P}=8.9e-6, broadband: R=0.86, \textit{P}=1.1e-6), indicating that channel capacity serves as an effective indicator for evaluating VEP BCI decoding performance at both the paradigm and individual level.

To gain a comprehensive understanding of the characteristics of the two BCI paradigms, we conducted spatial, spectral, and temporal analyses on the evoked response, as shown in Fig. 3 e-h. The spatial pattern of activation for both paradigms exhibited similarities, mainly localized in the occipital and occipital-parietal lobes. However, there were notable differences in the spectral and temporal characteristics. In the narrowband stimulation paradigm, the evoked power was primarily concentrated in the stimulation frequency and its harmonics. On the other hand, the broadband stimulation paradigm resulted in evoked power that spread across the entire spectral range, as depicted in Fig. 3f. By examining temporal waveforms of two classes, we observed that the correlation coefficient matrix between the classes revealed superior interference resistance in the broadband paradigm compared to the narrowband paradigm. This finding implies that the broadband response is less susceptible to interfere with other targets, resulting in a more even distribution of error probabilities, which provides an explanation for the observed performance enhancement in the broadband paradigm. Overall, the spatial, spectral, and temporal analyses shed light on the distinct characteristics of the two BCI paradigms and further support the advantages of the broadband paradigm in terms of interference resilience and information transfer.

Finally, the broadband visual BCI paradigm is been validated through online cued and free spelling experiments (\textbf{\textit{n}}=8), as in Table 1. The discrimination model undergoes training on 8 blocks of calibration data and is then tested on 5 blocks with online feedback. Subsequently, participants are prompted to type the phrase ‘faster than ever,’ encompassing 16 characters, spaces included, without visual cues. During the free spelling phase, the gaze shifting time is adjusted to 1 s to allow subjects to locate the next character. The average ITR* for both cued and free spelling reached 17.01 and 16.86 bps, respectively. These results underscore how the proposed WN paradigm can significantly contribute to the development of high-speed online BCIs.  

\begin{table}[]
\centering
\caption{Online experiment results}
\label{tab:my-table}
\begin{tabular}{ll|lll|lll}
\hline
\multicolumn{2}{l|}{} & \multicolumn{3}{c|}{Cued Spelling} & \multicolumn{3}{c}{Free Spelling} \\
Subject & Time (s) & Acc & ITR (bpm) & ITR*(bps) & Acc & ITR (bpm) & ITR*(bps) \\ \hline
S1 & 0.2 & 97\% & 426.1 & 24.86 & 91\% & 221.98 & 22.2 \\
S2 & 0.3 & 91\% & 335.26 & 14.9 & 95\% & 216.02 & 15.6 \\
S3 & 0.3 & 98\% & 377.35 & 16.77 & 94\% & 216.02 & 15.6 \\
S4 & 0.3 & 95\% & 360.82 & 16.04 & 94\% & 217.73 & 15.72 \\
S5 & 0.3 & 97\% & 369.09 & 16.4 & 95\% & 219.21 & 15.9 \\
S6 & 0.3 & 92\% & 337.89 & 15.02 & 94\% & 217.73 & 15.72 \\
S7 & 0.3 & 92\% & 343.82 & 15.28 & 99\% & 239.46 & 17.29 \\
S8 & 0.3 & 98\% & 378.11 & 16.81 & 98\% & 233.3 & 16.85 \\
\multicolumn{2}{c|}{Summary} & 95\% & 366.05 & 17.01 & 95\% & 222.68 & 16.86 \\ \hline
\end{tabular}%
\end{table}

\subsection{Analysis from information theoretical framework}
\begin{figure}
    \centering
    \includegraphics[width=\textwidth]{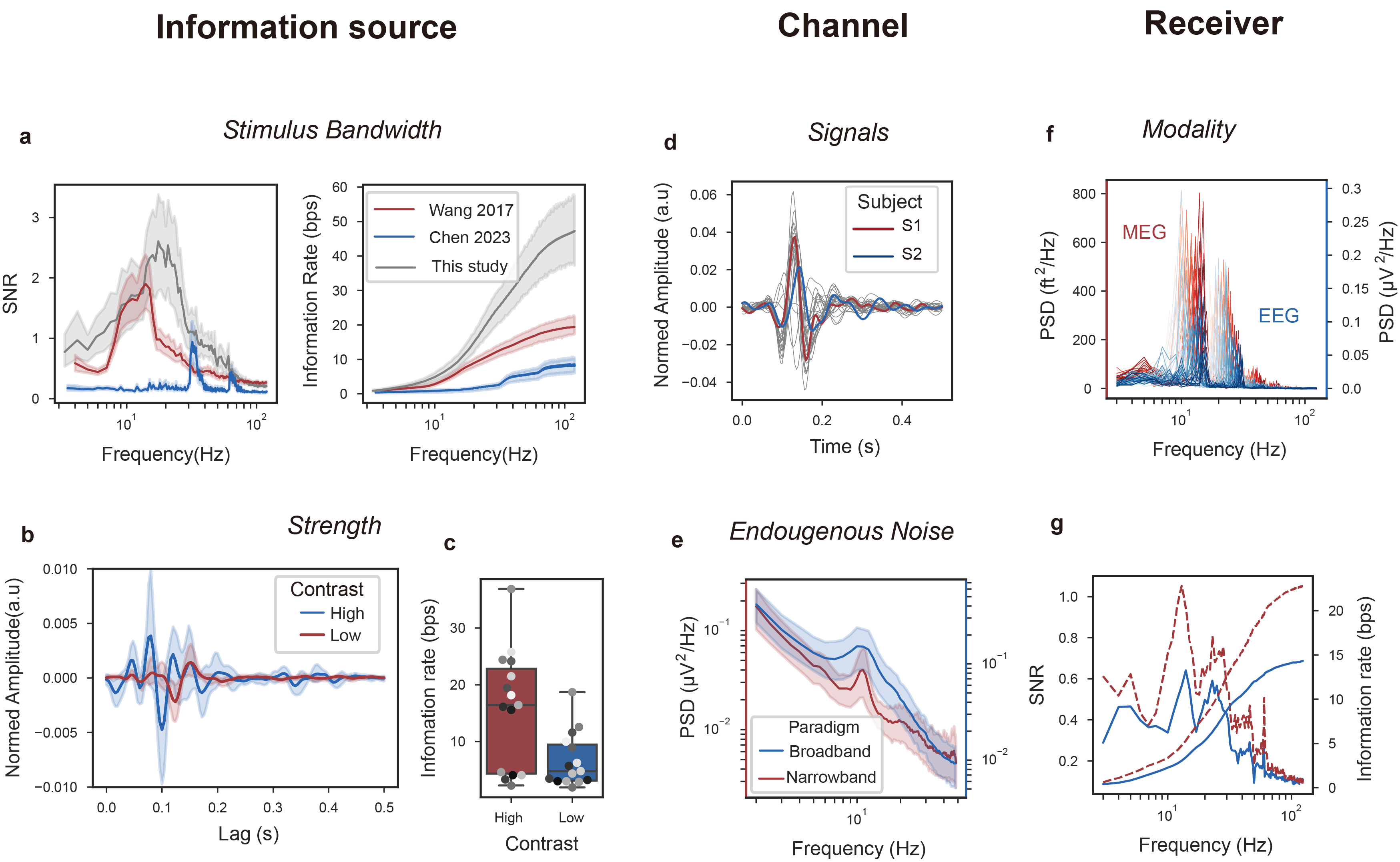}
    \caption{Analysis of related dataset under information theory framework. \textbf{a}, The increasing capacity estimated by different stimulus band width (gray: \textbf{\textit{n}}=20, red: \textbf{\textit{n}}=35\cite{wang2016benchmark}, blue: \textbf{\textit{n}}=20\cite{chen2023optimizing}, mean, 95\% CI)   \textbf{b}, The TRF under high and low contrast narrowband stimulation (sweep-1 dataset, \textbf{\textit{n}}=10, mean, 95\% CI). \textbf{c}, The resulting upper bound mutual information (\textbf{\textit{n}}=10, medians, interquartile ranges, maxima and minima). \textbf{d}, The TRF $h(\tau)$ for 20 subjects in this study, each line represents a subject. \textbf{e}, The spectral representation of noise component for two datasets (mean,95\% CI, blue: this study, preliminary experiment, \textbf{\textit{n}}=10; red: Wang et al.\cite{wang2016benchmark}, \textbf{\textit{n}}=35) \textbf{f}, The averaged spectral representation of SSVEP evoked response under simultaneously M/EEG, (mean, \textbf{\textit{n}}=5). \textbf{g}, The averaged upper bound SNR(f) and mutual information of M/EEG (\textbf{\textit{n}}=5)}
    \label{fig:enter-label}
\end{figure}
The progress in enhancing information rates has been achieved by fully exploiting the spectrum resources from the information source perspective. When analyzed within the information theoretical framework, it's essential to emphasize that, beyond the information source, the information channel and receiver also require scrutiny. For instance, it's evident that the spectrum resources of the primary visual system are inherently limited. Furthermore, whether the information can be fully captured noninvasively depends on the receiver, particularly the signal modalities involved. So, as we have demonstrated the importance of stimulus bandwidth, it is also crucial to note that several other factors can influence mutual information. Next, we aim to systematically inspect some of these factors through the lens of information source, channel, and receiver.

From the perspective of the information source, the status of stimulus paradigm inevitably affects the final information process. In this study, we used a stimulus bandwidth within 30 Hz to estimate the channel capacity, which might result in a slight underestimation. Thus, to demonstrate the spectral resource of other ranges, we first extended the analysis of stimulus bandwidth to two other published datasets. We include two SSVEP datasets covering two distinct frequency ranges (8-15.8 Hz\cite{wang2016benchmark}, and 31-41 Hz\cite{chen2023optimizing}). The SNR(f) of three datasets confirmed that spectral resources mainly concentrate in alpha and beta band. When the stimulation bandwidth is limited, the responses are primarily elicited within the stimulation band and its harmonics, thereby only utilizing a portion of the available spectrum resources (Fig. 4a). Notably, when the stimulation is located in the gamma band ($>$30 Hz), the information rate can be significantly decreased. In addition to bandwidth, stimulation strength, such as contrast, also has a substantial impact on the information rate. We analyzed a sweep SSVEP dataset covering frequencies from 1-60 Hz (\textit{Sweep 1 in Tabel 3}), acquired under low (50\%) and high (300\%) contrast conditions. The temporal dynamics $h(\tau)$ of the low contrast level is not only weaker, delayed, but also insensitive to high frequency band stimulation. Consequently, the results in Fig 4b implied that low contrast stimulation evoked much less spectrum resources than the high contrast, yielded in a significantly lower information rate (Fig. 4c). These findings align with previous studies\cite{stromeyer2003human}, which have also indicated the importance of selecting an optimal visual paradigm to achieve optimal information transfer in the primary visual channel. Specifically, the visual paradigm should cover the frequency bands of interest, such as the alpha and beta bands, while avoiding stimuli with low contrast.

No matter how much information we wish to flow from the information source, the information processing ability of the sensory system is limited\cite{donner2021temporal}. From the channel aspect, we can analyze both from the signal and noise components. The frequency preference of primary visual channel determines that alpha and beta band contribute greatly to spectrum resource. This spectral preference, modelled by $h(\tau)$, can vary across individual channels. Fig. 4d presents the individual variability by highlighting two subjects (red and blue; each gray line represents a participant) for comparison. Shaper and stronger $h(\tau)$ lead to rapider temporal dynamic response, indicating more information from the source can be filtered as signal component. Besides the signal, the characteristics of the noise component within the primary visual channel also play a significant role. In the proposed channel model, the noise can be mechanical or more relevantly, derive from endogenous neural activities. Our investigations reveal that in addition to the $1/f$ distributed color noise, there is a persistent endogenous oscillation at 10 Hz during both narrowband and broadband stimulation (Fig. 4e). Remarkably, the eigen oscillation persists across various stimulation conditions but can exhibit variations across individuals and possibly brain states (See Figure. S7 in Supplementary material).

Finally, from the information receiver’s perspective, the signal modality inevitably affects the estimated capacity and decoded ITR*. Many spectrum resources could be lost due to scalp attenuation. While EEG is frequently used in BCI studies, magnetoencephalography (MEG) offers distinct advantages due to its reduced scalp attenuation\cite{mellinger2007meg}. We confirmed this superiority using a simultaneous M/EEG (\textbf{\textit{n}}=5) dataset acquired under JFPM stimulation (Fig. 4f). The results clearly demonstrated that MEG was capable of capturing neural responses with a stronger SNR(f) compared to EEG (Fig. 4g), thereby leading to higher information rates. These findings highlight the potential for further enhancement in information transfer by leveraging intracranial recordings, where the proximity to the neural sources and the reduced signal attenuation through scalp and skull can exploit even greater spectrum resources and consequently decode higher information rates.

Although we focused on analyzing the primary visual channel, the principles of information theory are applicable to all BCIs that utilize sensory evoked channels for information transfer\cite{gao2014visual,lopez2012auditory}. The framework guides us to comprehensively consider the spectrum resources evoked by stimuli, the inherently limited capacity, and which modalities can capture information, thus aiding in understanding the decoded information rate of BCIs.
\section{Discussion}
While most research efforts have been focused on enhancing the information rate of BCIs through the SSVEP paradigm, we propose that the next breakthrough lies in the development of broadband code modulated VEP BCIs (cVEP BCI). This advancement is achievable by leveraging the information theory framework, which enables a deeper understanding of general information transfer in visual BCIs. Through this framework, we have discovered that stimulus design should prioritize optimal utilization of the spectrum resources of the human sensory system. 

\begin{figure}
    \centering
    \includegraphics[width=\textwidth]{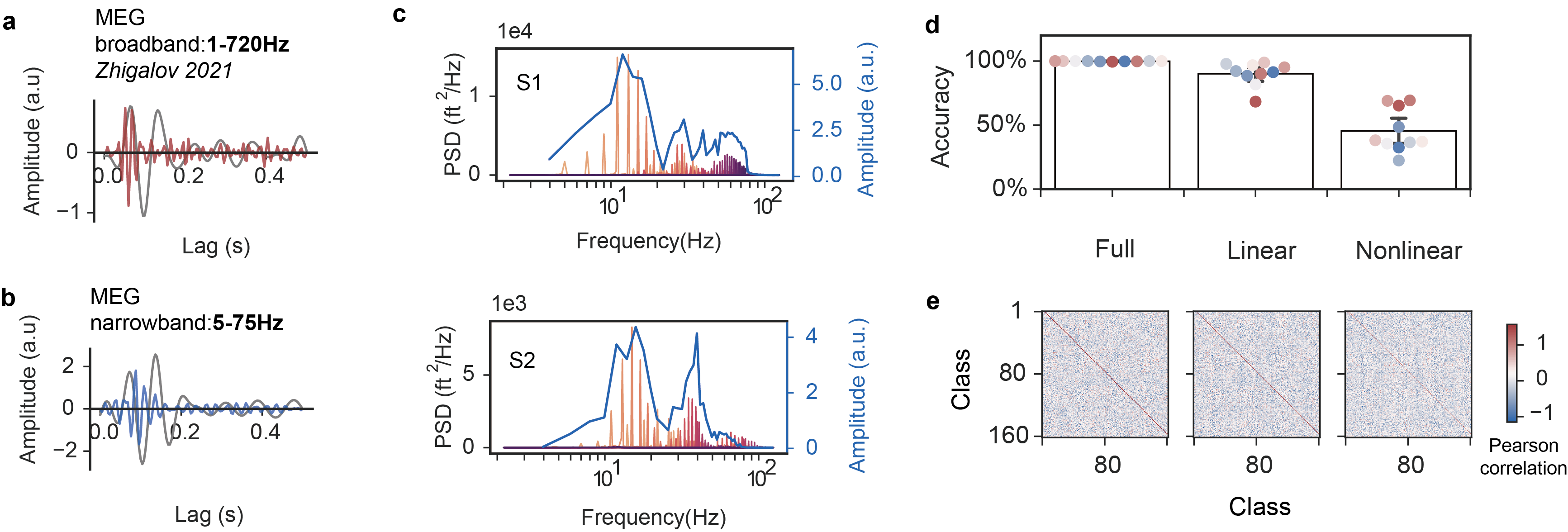}
    \caption{The alpha band (filtered to 8-20 Hz, background gray line) and gamma band (filtered to 20-100Hz, red line) TRF $h(\tau)$ of a representative participant, replicated from Zhigalov et al.\cite{zhigalov2021visual} \textbf{b}, the same as \textbf{c}, calculated from sweep-2 MEG (unpublish) with blue line represents the gamma component of TRF $h(\tau)$. \textbf{c}, The frequency response $H(f)$ (right blue y-axis) and the spectral representation of evoked response (each peak represents a stimulation frequency, 6-75 Hz) \textbf{d}, Upper panel, the classification accuracy based on full, linear and linear temporal templates (preliminary experiment data, \textbf{\textit{n}}=10, mean, error bar is 95\% CI), the linear templates are obtained through linear encoder. \textbf{e}, The Pearson coefficient matrices.}
    \label{fig:enter-label}
\end{figure}
\subsection{Use spectrum resources of primary visual channel}

The capacity of the human visual system to receive flicker stimulation if limited. This capacity, described by the SNR(f), represents the spectrum resources of primary visual channel. When visual BCIs allocate each target as a unique flicker sequence, they employ a multiple access approach to utilize the spectrum resources. SSVEP BCIs splits the spectrum resources to the non-overlapping sub-channels, which is referred to as Frequency-division multiple access (FDMA). First, the JFPM failed to stimulate the full system bandwidth, which certainly extend beyond 8-15.8 Hz. Secondly, narrowband identification is prone to interference from endogenous noise at 10 Hz. And most importantly, unlike ordinary communication system, primary visual channel exhibit strong a resonance property at 10 Hz, which means the spectrum resource is distributed unevenly across the frequency range. This could result in varying error rates for targets allocated at different frequencies. On the contrary, targets in the broadband BCIs share the spectrum resources over the whole frequency range, which is referred to as Code-division multiple access (CDMA). The CDMA employs the spread spectrum technique and thus gains several advantages over FDMA\cite{zigangirov2004theory}. It’s less likely to experience interference with narrowband noise. It’s easy to generate a large number of targets that evenly utilize the complete spectrum resources, paving the way to build visual BCIs with more than a hundred targets. Consequently, broadband BCI demonstrates a more efficient utilization of the spectrum resources of the primary visual channel. 

Previous studies have also proposed broadband BCIs based on m-sequences
\cite{bin2011high}; however, they did not achieve information rates comparable to JFPM SSVEP BCIs. The main reason for this is that traditional m-sequences are inflexible to generate for any given length, and the binary coding leads to severe visual discomfort. Recently, a few other studies have explored alternative approaches to address the limitations of traditional broadband BCIs. For example, Martínez et al.. introduced the use of multi-level m-sequences instead of binary sequences\cite{martinez2023non}, which enhanced visual comfortability during stimulation. Nagel et al.. suggested the use of arbitrary binary codes instead of m-sequences\cite{nagel2019world}, which improve the flexibility of broadband coding. And recently, Xu et al. investigated the influence of refresh rate on stimulus bandwidth\cite{xu2023stimulus}. Compares to these studies, the WN code stands out by offering an alternative broadband coding scheme that not only provides flexibility and visual comfort but also enhances our understanding of information transfer in the primary visual channel. 

The next crucial question is: Is there a broader range of spectrum resources that we have yet to tap into? Accumulated evidence suggests that the spectrum resource primarily resides in the alpha and beta frequency ranges\cite{fawcett2004temporal,gulbinaite2019attention,zhigalov2021visual,groen2022temporal}. However, it remains uncertain if the spectrum resource extends to higher gamma bands and if it can be utilized noninvasively. In relation to this question, Gulbinaite et al. identified a gamma resonance peak around 40 Hz in the spectral domain using narrowband stimulation\cite{gulbinaite2019attention}. More recently, Zhigalov et al.. reported coherent MEG results utilizing broadband white noise stimulus (Fig. 5a), revealing the presence of not only alpha but also gamma band components (gamma echo) in the temporal domain $h(\tau)$\cite{zhigalov2021visual}. From a linear system perspective, the components in the temporal domain $h(\tau)$ must correspond to resonance peaks in the spectral domain $H(f)$\cite{donner2021temporal}.In our study, we confirmed the presence of the gamma component in both $h(\tau)$ and $H(f)$ using Steady State Visual Evoked Field (SSVEF) sweep data (\textbf{\textit{n}}=9) ranging from 5 to 75 Hz with a 1 Hz interval. We concluded that the gamma band resonance peak does exist (Fig. 5b), although it is weaker than the alpha band peaks and may vary considerably among participants (Fig. 5c, Figure S6 in Supplementary material). While exploiting the gamma band resonance peak may lead to a marginal increase in the information rate, further investigation is necessary to fully understand its potential.

\subsection{Nonlinear modeling and decoded information}

The difference between the upper and lower bound information rates reflects the efficiency of the encoding model, and thus related to different category of decoding algorithms. For template matching algorithms like TDCA and TRCA, which utilize individual averaged responses as matching templates, the decoded information should be evaluated against the upper bound limit. Conversely, for training-free algorithms relying on simulated responses, comparison should be made with the lower bound, as they often incorporate reconstructed responses as matching templates. In this study, the signal component of the lower bound was modeled by a linear convolution process using the $h(\tau)$. However, it is important to note that the linear encoder neglects the nonlinear components present in the actual neural responses. To quantify these linear and nonlinear components, we classified 160 classes of broadband responses by analyzing the full evoked response, the linearly reconstructed response, and the nonlinear response derived by subtracting the reconstructed temporal templates from the real data. The results in Fig. 5d indicated that while the accuracy of the nonlinear responses was lower compared to the linear reconstructed responses, they still achieved an accuracy of 45\% (90\% for the linear). The results indicated that there is still great amount of information waiting to be exploit in the nonlinear components. These nonlinear components can be break down to light adaptation, distinct neural dynamics under high/low contrast, and refractory response under short Inter-Stimulus Interval (ISI), etc\cite{groen2022temporal,zhou2019predicting}. Moving forward, future studies should further develop these linear-nonlinear models to enhance our understanding of noninvasive temporal dynamics. With the aid of nonlinear encoding, we expect to estimate more convergent lower and upper bounds\cite{butts2007temporal,butts2011temporal,pillow2008spatio}. This advancement could significantly support the development of high-performance broadband BCIs in a training-free manner.

\subsection{Limitations of channel hypotheses}

As we expand the theoretical framework of channel capacity to noninvasive visual BCIs, we inherit certain hypotheses from previous studies that can impact the estimation. One significant hypothesis is the additive Gaussian channel assumption, which suggests that the signal component can be obtained through trial averaging, while the noise is obtained through residual subtraction. However, it is important to note that in reality, this hypothesis may not always be applicable, as induced neural responses can be canceled out after trial averaging\cite{david2006mechanisms}. Therefore, the analysis conducted in this study is primarily focused on contrast modulated responses, where the signal component is typically considered to be evoked. It is important to note that various stimulus types, such as moving gratings\cite{swettenham2009spectral} or natural images\cite{lesica2007adaptation}, can activate distinct channels within the neural system, potentially surpassing the capacity of the contrast-modulated primary visual channel as reported in our study.

Beyond the hypotheses inherited from previous studies, our approach involves the utilization of a Single-Input Single-Output (SISO) system for multi-channel noninvasive recordings. However, it's essential to acknowledge that the SISO hypothesis simplifies neural dynamics to a single principal component, potentially leading to an underestimation of the true information rate. Various studies have highlighted that additional components could hold independent information as well\cite{zhigalov2023perceptual}. To address this limitation, forthcoming research should consider integrating source imaging techniques to identify valid independent components, thus providing a more precise estimation of the channel capacity.

\section{Conclusion}

After years of observing limited progress in the advancement of information rates, we have achieved a breakthrough by setting a new record for decoded ITR using a paradigm contrary to narrowband SSVEP—specifically, the broadband paradigm.This growth can be attributed to the enhanced mutual information achieved by expanding the stimulus bandwidth. This study not only introduces a new direction for VEP BCI research centered around broadband modulation but also presents a universally applicable theoretical framework for examining information flow in all sensory-evoked BCIs. We hope that this contribution will pave the way for the development of practical and usable non-invasive BCIs, ushering in the next generation of communication systems.

\section{Method}
\subsection{Generate dynamic stimulus}

This study studied broadband temporal dynamic stimulus for comparison with narrowband SSVEP stimulus. The broadband stimulus is the uniform-distributed random white noise (WN). Unlike traditional m-sequences generated from shift registers, the WN sequence can be generated at any time window, covering a multi-level value range from 0 to 1. The randomness of the sequences is controlled by setting a seed, ensuring that the sequences are generated consistently and repeatedly. The narrowband stimulus follows the JFPM paradigm, which encodes each stimulation class as sinusoidal waveforms:

\begin{equation}
\left.x_n(t)=\sin \left\{2 \pi\left[f_0+(n-1) \Delta f\right] t+\left[\phi+(n-1) \Delta \phi\right)\right]\right\}, n=1, \cdots, N
\end{equation}
\begin{figure}[!htbp]
    \centering
    \includegraphics[width=\textwidth]{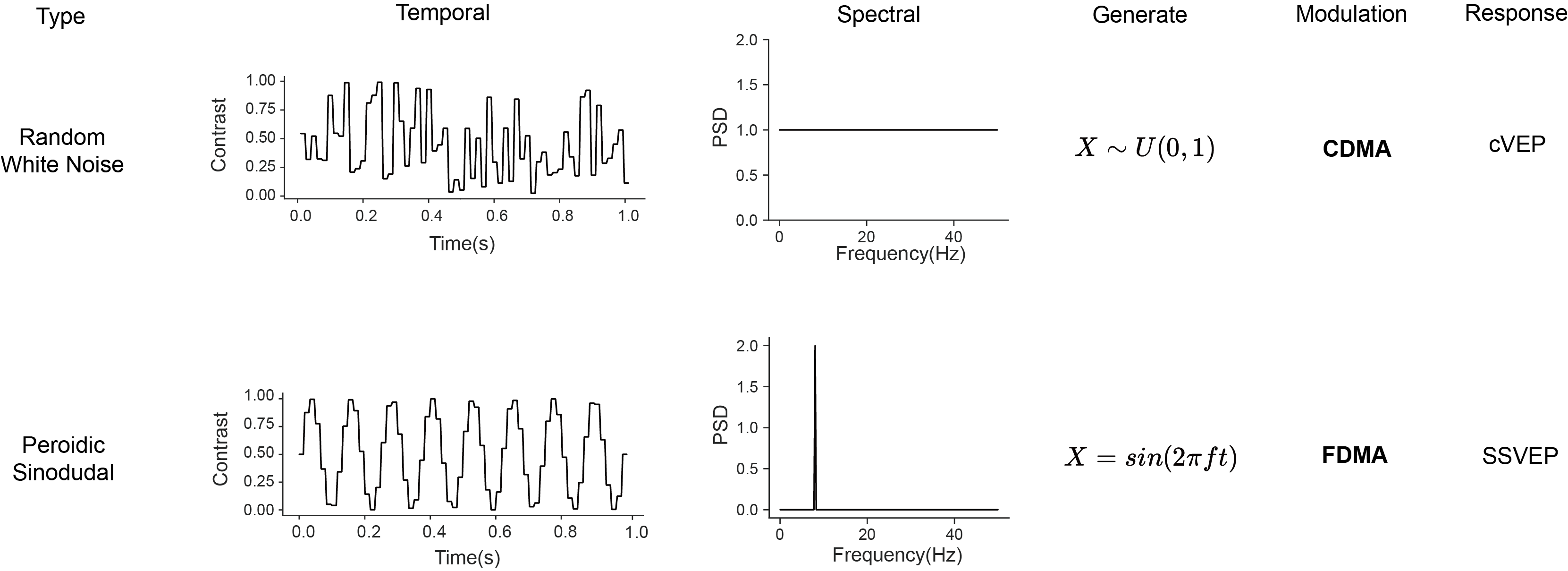}
    \label{fig:enter-label}
\end{figure}
, where $f_0$ and $\phi_0$ denotes the initial frequency and phase, the $\Delta f$ and $\Delta \phi$ is the interval of adjacent target. Here in this study, we set $f_0$=8 Hz, $\phi_0=0 \pi$, $\Delta f$=0.2 Hz, $\Delta \phi$=0.35 $\pi$. In contrast to manually setting the bandwidth to 8-15.8 Hz, the bandwidth of the WN sequence is constrained by the Nyquist frequency of the screen refresh rate (60 Hz in this study), resulting in a 1-30 Hz bandwidth. Within this range, the flexibility of WN coding allows us to generate a large number of targets to meet experimental requirements. For instance, in this study, we generated up to 160 classes of sequences. However, to provide a fair comparison with the state-of-the-art 40-target SSVEP speller design, we also need to design a 40-target broadband WN speller for comparison (detailed in the following optimization section).

\subsection{Information theoretic framework and linear modeling }

To properly estimate the mutual information of the primary visual channel, we adopt the theoretic framework reviewed in related\cite{borst1999information}. Here, the primary visual pathway is modeled as a gaussian channel with additive noise, which characterizes the pathway that conveys information from stimulus $S$ to response $R$ with the addition of independent Gaussian noise $N$. 
\begin{equation}
\mathrm{R}=\mathrm{S}+\mathrm{N}
\end{equation}

In such framework, the information contained in stimulus and response is represented as I(S,R), which can be obtained by Signal-to-Noise Ratio (SNR) in frequency domain:
\begin{equation}
I(S, R)=\int_0^{f 1} \log _2[1+S N R(f)] d f
\end{equation}
, where the $f1$ states the integral limits.

By adopting different hypotheses on the encoding model, we can devise two distinct approaches to compute SNR(f), resulting in two bounds on the theoretical limits: the upper and lower bound  . The upper bound method estimates from the response domain and do not make any assumptions on the stimulus-response functions. To calculate the upper bound, the same dynamic stimulus is presented multiple times to elicit the corresponding response. This response is then averaged across n trials to obtain an estimate of the average response, which serves as the signal component. The difference between the stimulus and response is considered to be noise to noise. The difference between $S$ and $R$ is thought to be noise. 

\begin{equation}
N_{\text {upper }}=R_i-\sum_i R
\end{equation}

The essential assumption of the upper bound method is that the evoked averaged neural response contained all the encoded information, while the background noise is independent of the evoked response. Such hypothesis is widely used, especially for non-invasive studies. Thus, the SNR can be calculated as:
\begin{equation}
SNR_{\text {upper}}(f)=\frac{|\bar{R}(f)|^2}{|N(f)|^2}
\end{equation}
Where $\bar{R}(f)$ and $N(f)$ represent the spectral representation of signal and noise. From communication perspective, the upper bound information is called channel capacity.

Unlike the upper bound, the lower bound is estimated from stimulus perspectives and post an assumption on the stimulus-response function, namely the encoding model. We used linear encoder in this study, which hold the view that neural response can be acquired through convolving the impulse response function $h(\tau)$ with the dynamic stimulus (See Figure S1 in Supplementary material). 

\begin{equation}
R(t)=\sum h(\tau) S(t-\tau)+\varepsilon(t)
\end{equation}
The $h(\tau)$  can be obtained by:
\begin{equation}
h(\tau)=\left(\boldsymbol{S}^{\mathrm{T}} \boldsymbol{S}\right)^{-1}\left(\boldsymbol{S}^T \cdot R\right)
\end{equation}
, where S is the Toeplitz matrix of stimulus pattern. By modeling the pathway as a linear encoder, we see the primary visual pathway a bandpass filter $H(f)$ in the frequency domain. 
\begin{equation}
S_{\text {est }}(f)=H(f) \cdot R(f)
\end{equation}

In this sense, we can reconstruct the stimulus from neural response to measure the information received. Thus, the signal in the lower bound is the signal trial reconstruction of S, and the difference is set to be the noise:
\begin{equation}
N_{\text {upper}}=R_i-\sum_i R
\end{equation}
Finally, the SNR(f) of low bound can be calculated as:

\begin{equation}
S N R_{l b}(f)=\frac{\mathrm{S}_{\mathrm{est}}(\mathrm{f}) \mathrm{S}_{\mathrm{est}}^*(f)}{N(f) N^*(f)}
\end{equation}
where *  represents the complex conjugate. From linear modelling perspective, $h(\tau)$ or$H(f)$ fully characterized the spectral distribution of the studied channel, and naturally determined the information it transferred. When estimating SNR across multiple stimulus class and trials, caution must be exercised to avoid overestimation. Statistically, the final SNR should be adjusted as:

\begin{equation}
S N R_{l b}(f)=\frac{\mathrm{S}_{\mathrm{est}}(\mathrm{f}) \mathrm{S}_{\mathrm{est}}^*(f)}{N(f) N^*(f)}
\end{equation}
where m stands for the trial number of each class, and the $<>$ stands for average across stimulus classes\cite{passaglia2004information}.

Unlike single neuron recordings, the noninvasive field potentials are acquired in multi-channel setting. To simply the information processing process, we use spatial filters to extract source component from multivariate recordings. This simplification assumes a Single Input and Single Output (SISO) channel model, which is reasonable as the luminous modulation response typically contributes to one or a few latent components in the source space. It's important to note that the mutual information estimation can be influenced by the choice of spatial filter algorithms. We chose the TDCA algorithm to construct the spatial filter\cite{liu2021improving}, which is consider to be one of most effective ones for VEP BCI.

To set information transfer in the BCI context, researchers have come up with the formula that also derive from the Shannon theorem:

\begin{equation}
I T R=60 \cdot\left(\log _2 M+P \log _2 P+(1-P) \log _2 \frac{1-P}{M-1}\right) / T
\end{equation}
, where M, P, and T denotes the number of stimulus class, accuracy, and decoding time, respectively. Particularly, in BCI researches, a gaze shift time usually ranges from 0.5 – 1s are considered in the decoding time. When considering the theoretical information transfer, the gaze shift time (0.5 s in this study) should be excluded since during this time, no stimulus was received by the channel. In this study, we denote this theoretical ITR as ITR* in bps, which is the same as I(S,R).

\subsection{Decoding algorithm}

We chose the current best Task Discriminate Component Analysis (TDCA) as the target identification algorithm\cite{liu2021improving}. The TDCA seeks a common projection weight $W^T$ to boost discriminability of multi-channel VEP based on the Fisher criterion. The Fisher criterion seeks a linear projection $W^T$ to maximize of the ratio osf between-class scatter to the within-class scatter. Consider the $i_{th}$ trial of calibration data $\boldsymbol{X}^{(i)}\in \mathbb{R}^{N_{ch}\times N_{s}} ,i=1,2,···,N_t$, where $N_{ch}$ is the number of channel and $N_s$ is the number of sample points. We can compute the within and between scatter matrices $\boldsymbol{H_w}$,$\boldsymbol{H_b}$ according to:
\begin{equation}
\begin{gathered}
\boldsymbol{H_b}=\frac{1}{\sqrt{N_c}} \cdot\left[\overline{\boldsymbol{X}}^{\mathbf{1}}-\overline{\boldsymbol{X}}^{a l l}, \cdots, \overline{\boldsymbol{X}}^{N_c}-\overline{\boldsymbol{X}}^{\text {all }}\right] \\
\boldsymbol{H_w}=\frac{1}{\sqrt{N_t}}\left[\boldsymbol{X}^{(\mathbf{1})}-\overline{\boldsymbol{X}}^{(\mathbf{1})}, \cdots, \boldsymbol{X}^{\left(N_t\right)}-\overline{\boldsymbol{X}}^{\left(N_t\right)}\right]
\end{gathered}
\end{equation}
Based on the scatter matrices, the weight $W^T$ can be obtained by:
\begin{equation}
\operatorname{maximize}_W=\frac{\operatorname{tr}\left(W^T S_b W\right)}{\operatorname{tr}\left(W^T S_w W\right)}
\end{equation}
, where the matrices $S_b$ and $S_w$ is the covariance matrix by $ \boldsymbol{S_b}=\boldsymbol{H_b} \boldsymbol{H_b^T}$ , $\boldsymbol{S_w}=\boldsymbol{H_w} \boldsymbol{H_w^T}$ . Different from the original TDCA study, we remove the data argumentation and referential signal projection techniques, only to preserve the spatiotemporal filtering process.

\subsection{Broadband Stimulus Optimization}

Due to the random nature of WN coding, there is a theoretically infinite number of random code sets possible with an infinite stimulation time. Therefore, it becomes essential to optimize the code set for an efficient broadband BCI design. To address this, we propose a two-step optimization scheme that involves group optimization followed by personal optimization (See Figure S2 S3 in Supplementary material).

Firstly, we perform group optimization to identify random code combinations that are as separable as possible at the group level. To achieve this, our objective is to maximize the minimum Euclidean distance between the between-class responses within each code combination. Because it's impractical to acquire a large number of real responses, we chose to optimize based on estimated responses. In detail, we first collect the associated evoked responses $R_0$ from 10 individuals receiving WN stimulation of preliminary code set $C_0$(160 class, randomly picked)  . The number of class (n=160) is decided by leveraging the experiment time, further studies can choose any class number for optimization. Based on the preliminary data  $(C_0,R_0)$ ,we built the linear encoder model to estimate the group-level temporal dynamics $(\mathbb{C},\mathbb{R})$. The group level response R can be simulated at any given stimulus, which allow us to pick the group-optimized code set $\hat{C}$   quickly from the whole code set space based on the simulating Annealing algorithm. The code set $\hat{C}$ is the coarse estimate to ensure that the stimulus patterns are as distinct as possible. However, it may not necessarily perform optimally for each individual participant (Figure S2 in Supplementary material).

To enhance individual performance in the calibration phase, we proceed with personal optimization. We record the neural responses to the group-optimized code set $(\hat{C})$. From this data, we manually select the customized code set ($C_i^*$) consisting of 40 classes that yield the best performance. Evaluating the BCI performance of numerous randomly sampled code sets can be impractical due to the time-consuming nature of calculating spatial filters for different code sets. Fortunately, the TDCA algorithm significantly reduces the need for multiple spatial filters by providing a single discriminative common set of weights ($W^T$) that can be applied to any combination of WN sequences. Therefore, during the personal optimization process, we first calculate the weights ($W^T$) for the 160-class code set $(\hat{C})$ and then randomly sample 40 classes from Pearson's correlation confusion matrices. We compute the decoding accuracy for each sampled code set without the need for time-consuming computations of spatial filters. The ultimately selected code set ($C_i^*$) guarantees suitability for individuals while minimizing computational burden.

\subsection{Offline and online experiment}

The experiment consists of three phases: preliminary, offline comparison, and online speller experiments, which serves different purpose. The stimulation was displayed on a 1920*1080 LCD screen (Acer GD245 HQ) at 60 Hz refresh rate. The EEG was recorded using a Synamps2 system (NeuroScan, Inc.) with a sampling rate of 1000 Hz. In the preliminary and offline experiments, 62 channels were recorded. For the online spelling experiments, only occipital lobe electrodes were used (N=21, Pz, P1/2, P3/4, P5/6, P7/8, POz, PO3/4, PO5/6, PO7/8, Oz, O1/2, and CB1/2) were used in the online spelling experiments. All recordings were conducted in a soundproof room at a distance of approximately 65 cm from the screen. Contact impedance was maintained at 15 $k\Omega$ throughout the experiments. Participants with normal or corrected-to-normal vision were recruited and required to read and sign a consent form approved by the Research Ethics Committee of Tsinghua University.

\begin{table}[]
\centering
\caption{Experiment parameters}
\label{tab:my-table}
\begin{tabular}{ccccc}
\hline
Experiments & Codeset & Class Number & Participant & Time(s) \\ \hline
Preliminary & $C_0$ & 160 & 10 & 1 \\
\multirow{2}{*}{Offline} & $\hat{C}$ & \multirow{2}{*}{40} & \multirow{2}{*}{20} & \multirow{2}{*}{0.5} \\
 & $C^*_i$+JFPM &  &  &  \\
Online & $C^*_i$ & 40 & 8 & 0.2-0.3 \\ \hline
\end{tabular}%
\end{table}

The preliminary experiment received single target broadband stimulation was used to estimate the upper and lower bound of mutual information, and to render the group optimization code combination $\hat{C}$. 10 participants (5 female) participated these experiments. During this experiment, participants were instructed to gaze at a single target flicker (50*50 cm) at the center of the screen. Each block of recordings consists of 160 trials corresponding to the 160 WN sequence in $C_0$. Each trial lasts 1 second with a 0.5-second interval. Every eight trials, the participants were given the opportunity to rest voluntarily and resume at their own pace. A total of six blocks were recorded for each participant.

The offline experiment is to compare the BCI performance of narrowband JFPM paradigm with the proposed broadband WN paradigm. 20 participants (10 female) participated this part. First, each participant was required to gaze on the single target broadband flicker ($\hat{C}$) to determine their personal-optimized code set $C^*_i$. Afterwards, the BCI performance was compared between the WN BCI encoded by $C^*_i$ with the SSVEP BCI encoded by the JFPM. In offline experiment, each trial consists of 0.5 s of stimulation and a 0.5-interval. Finally, the participants engaged in both broadband and narrowband cued-spelling task for eight blocks. To mitigate the effects of visual fatigue, the order of the two paradigms was randomized and balanced among the participant groups. Each trial began with a red cue lasting 0.5 seconds, indicating the target to be attended. This was followed by stimulation lasting 0.5 seconds. The participants were instructed to shift their gaze with in the duration of visual cue as soon as possible and to avoid eye blink. During each block, participants are asked to rest voluntarily and resume after a few minutes.

The online spelling aims to test the online spelling performance of broadband BCI. We recruited 10 subjects from the offline experiment to perform both cued and free spelling task. The cued spelling followed the same procedure as offline experiments, whereas the stimulation time is adjusted to 0.2-0.3 s, with a 0.5 s interval. The cued spelling last eight blocks of training and five blocks of testing. The free spelling required the subject to spell the phrase “faster than ever” (16 characters including space) for 5 times without correction. Different from the cued task, the interval in free spelling task was set to be 1 s for participant to find the target character.

\subsection{Channel analysis from information source , channel, and receiver}
Besides the narrowband and broadband data acquired in this study. We incorporate several other datasets for information analysis. The included datasets consist of EEG or MEG data acquired under both broadband and narrowband visual stimulation conditions, with variations in stimulation bandwidth. Detailed descriptions of datasets from published studies can be found in the respective references. Apart from those, the Sweep 1 dataset comprises EEG responses to sinusoidal flicker ranging from 1-60 Hz with a 1 Hz interval. In this dataset, 10 subjects fixated on a single target in the central visual field while the visual stimulation was presented on an LCD screen. Similarly, the Sweep 2 dataset involves a similar sweep frequency experiment with 9 subjects, ranging from 7-75 Hz with a 1 Hz interval. The stimulation was displayed as a single target on a ProPixx projector. Additionally, we incorporate a concurrently recorded M/EEG dataset of 5 subjects who underwent classical JFPM stimulation, presented on the same ProPixx projector.

\begin{table}[!h]
\centering
\caption{Description of Datasets}
\label{tab:my-table}
\begin{tabular}{cccccc}
\hline
Study & Signal Modality & Paradigm & Stimulation Bandwidth & Subjects & Time (s) \\ \hline
This study & EEG & cVEP & 1-30 & 20 & 0.5 -1 \\
Wang 2018\cite{wang2016benchmark} & EEG & SSVEP & 8-15.8 & 35 & 5 \\
Chen 2023\cite{chen2023optimizing} & EEG & SSVEP & 31-46 & 21 & 2 \\
Zhigalov 2022\cite{zhigalov2021visual} & MEG & cVEP & 1-720 & 4 & 3 \\
Sweep1 & EEG & SSVEP & 1-60 & 10 & 5 \\
Sweep2 & MEG & SSVEP & 5-75 & 9 & 8 \\
M/EEG & M/EEG & SSVEP & 8-15.8 & 5 & 5 \\ \hline
\end{tabular}%
\end{table}

\newpage
\newpage

\section{Supplementary Material}

\setcounter{figure}{0}

\makeatletter 
\renewcommand{\thefigure}{S\@arabic\c@figure}
\makeatother

\begin{figure}[h]
    \centering
    \includegraphics{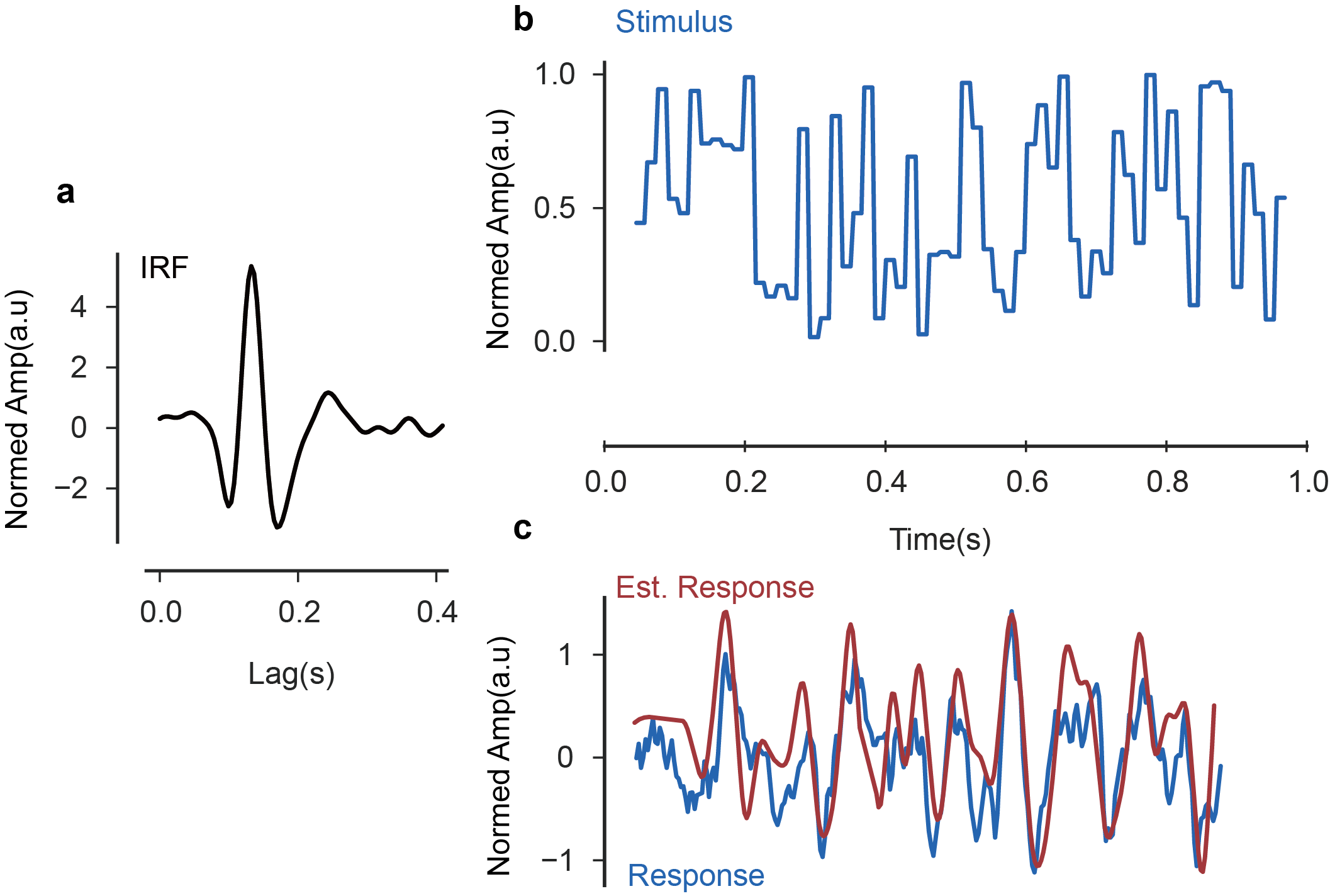}
    \caption{Estimated broadband response from linear encoder model. \textbf{a}, The TRF $h(\tau)$ from a representative subject. \textbf{b}, A class of WN stimulus. \textbf{c}, Blue: the actual evoked response corresponds to the stimulus above, red: estimated response by linear encoder model.}
    \label{fig:enter-label}
\end{figure}
\newpage

\begin{figure}[h]
    \centering
    \includegraphics[width=\textwidth]{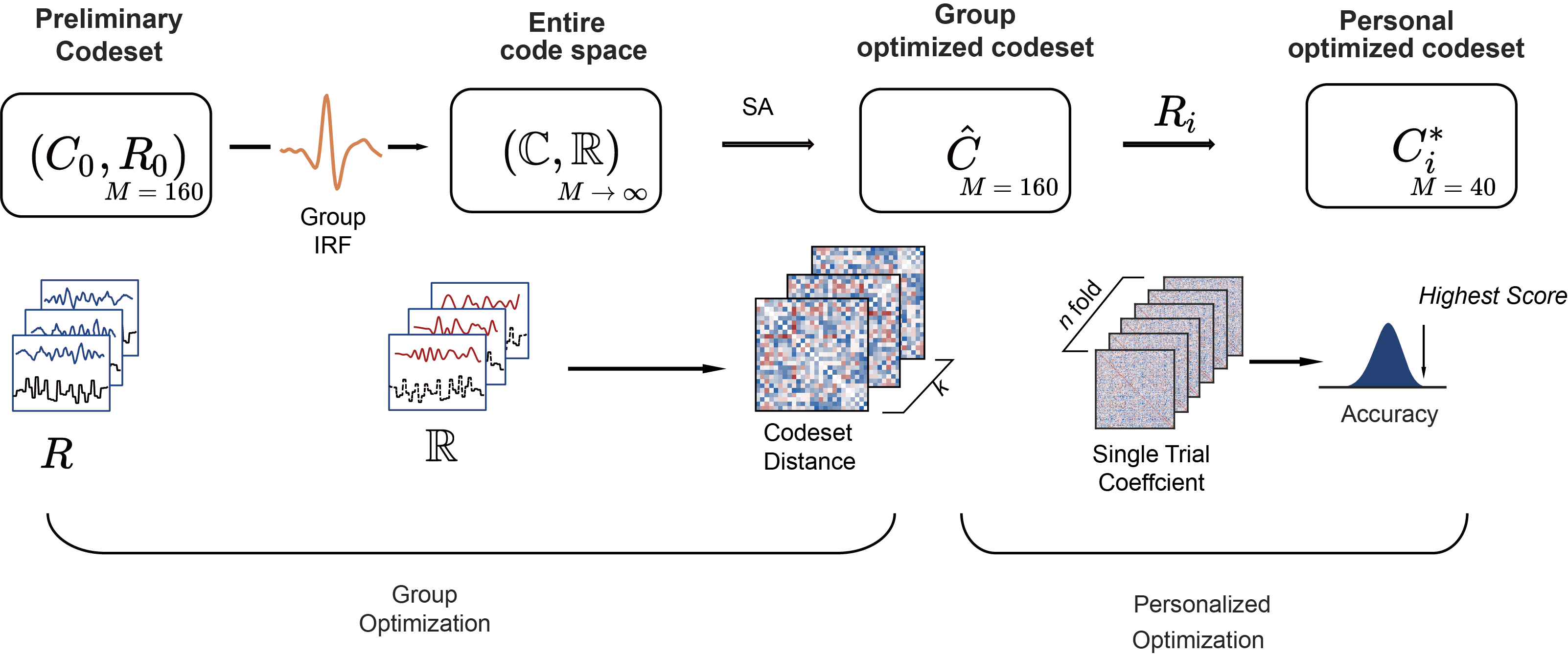}
    \caption{The optimization procedure. The group optimization is to maximum between-class separability measured by Euclidean distance. The Simulated Annealing (SA) algorithm to find the maximum code set distances between estimated group response. The personalized optimization is to directly select the best-performed combination by random sampling $10^4$ times.}
    \label{fig:enter-label}
\end{figure}
\newpage

\begin{figure}[h]
    \centering
    \includegraphics{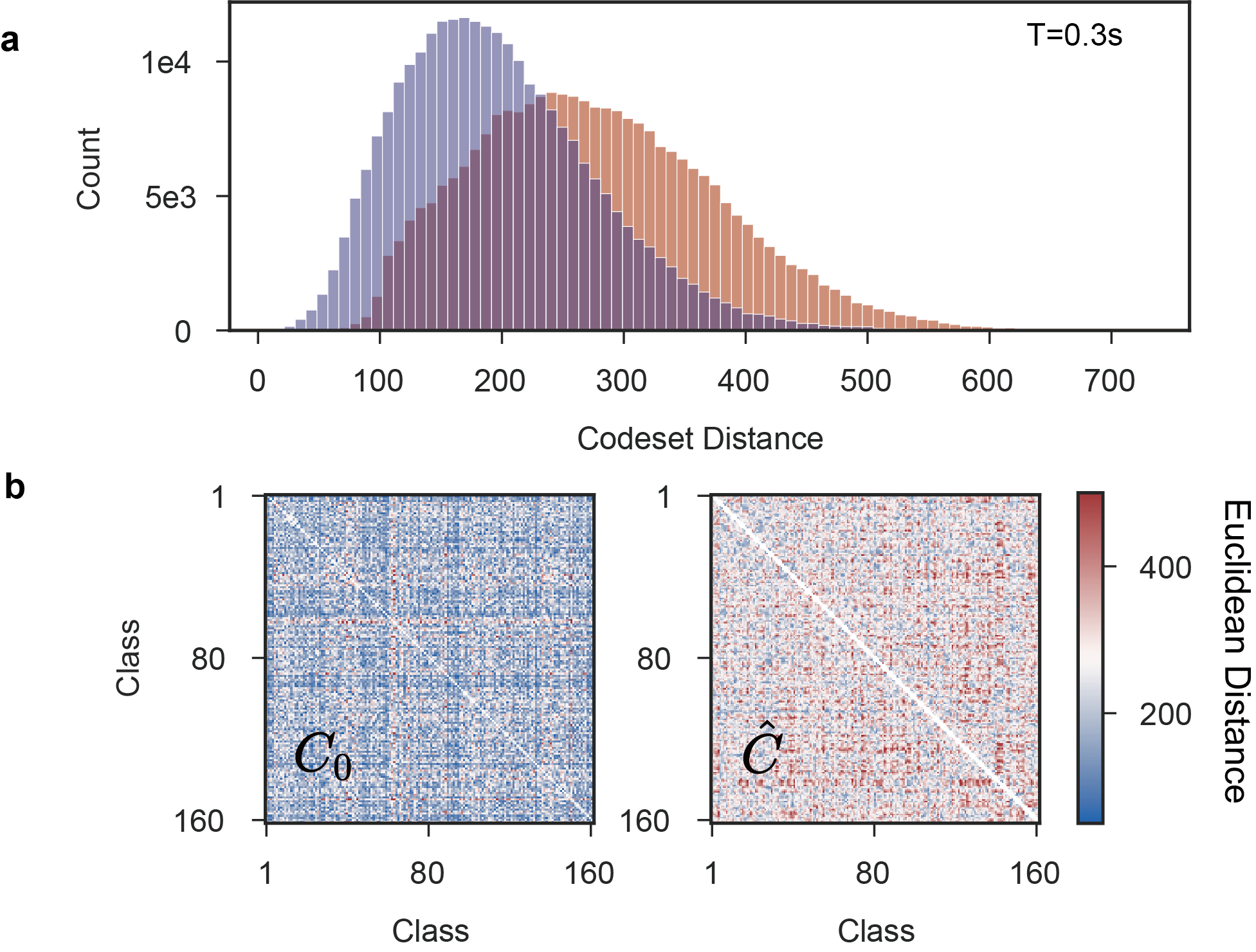}
    \caption{The between-class Euclidean distance before ($C_0$) and after ($\hat{C}$) SA optimization. a, The histogram of pairwise code set distance. b, the confusion matrices of $C_0$ and $\hat{C}$}
    \label{fig:enter-label}
\end{figure}

\newpage

\begin{figure}[h]
    \centering
    \includegraphics[width=\textwidth]{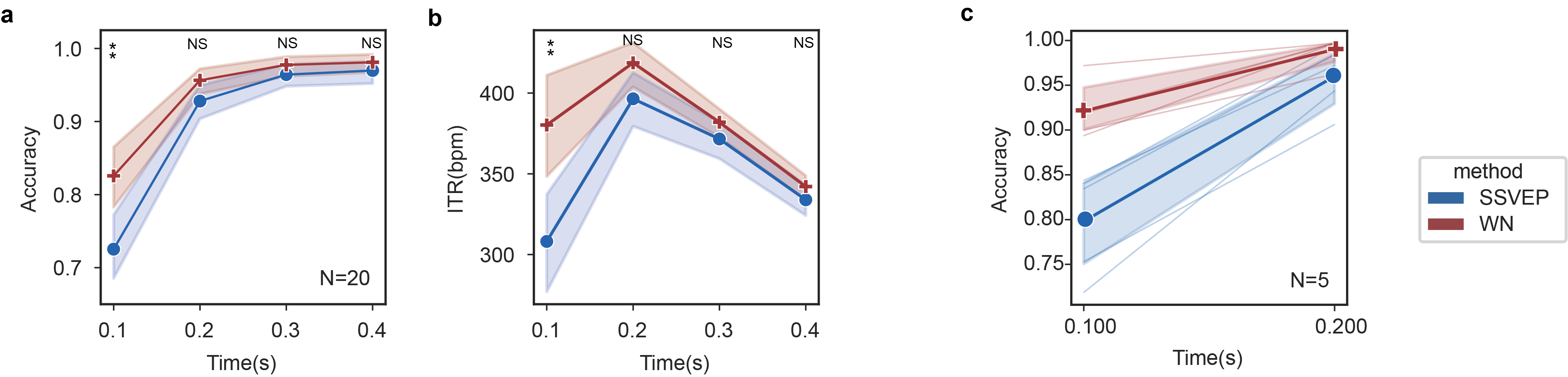}
    \caption{The BCI performance including ITR. \textbf{a}, left: the same as figure 2a, right: the ITR in bpm (\textbf{\textit{n}}=20, \textit{P}=0.002 at 0.1 s, paired t-Test, mean,95\% CI). \textbf{b}, The best performed 5 subjects under short target identification time. The bright solid lines denote the averaged accuracy, the background lighter lines as individual samples (mean, 95\% CI).}
    \label{fig:enter-label}
\end{figure}

\newpage

\begin{figure}[h]
    \centering
    \includegraphics[scale=0.9]{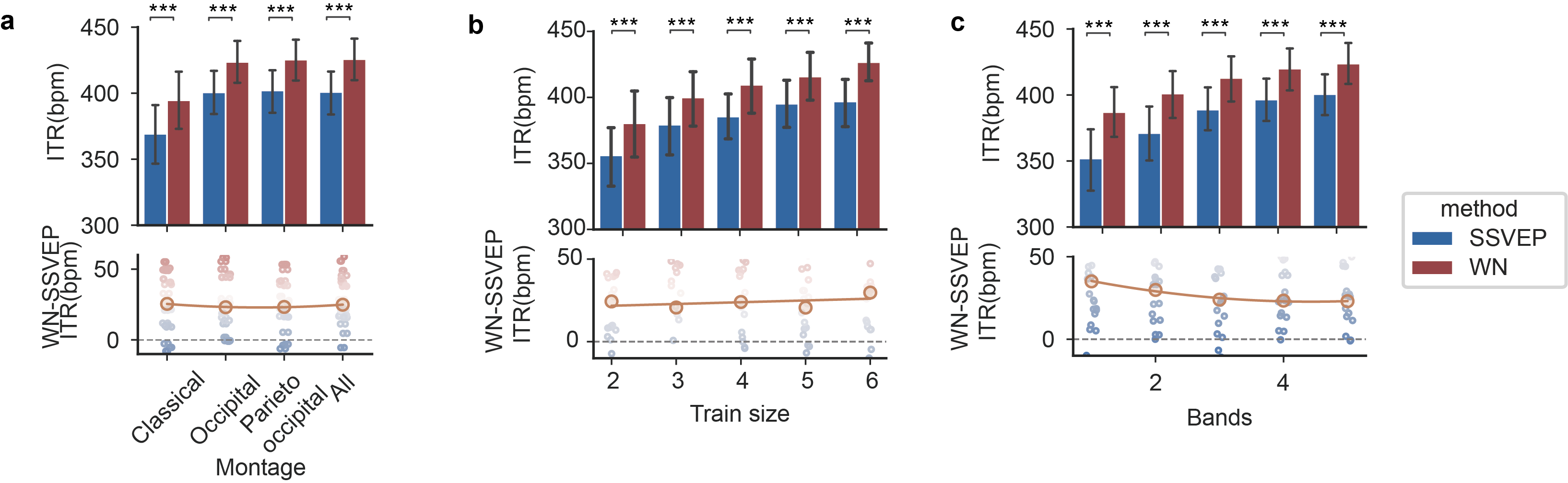}
    \caption{The highest ITR with the increase of train data. \textbf{a}, The impact of channel montages (\textbf{\textit{n}}=20, mean, error bar is 95\% CI, the lower panel is the difference between two paradigms, orange line is the mean, background dots represents each subjects. Paired t-Test, ***represents P<0.001.Classical:9 channels, Pz, POz, PO3/4, PO5/6, Oz and O1/2;Occipital:21 channels, Pz, P1/2, P3/4, P5/6, P7/8, POz, PO3/4, PO5/6, PO7/8, Oz, O1/2, and CB1/2; Parieto-occipital: 30 channels, CPz, CP1/2, CP3/4, CP5/6, TP7/8, Pz, P1/2, P3/4, P5/6, P7/8, POz, PO3/4, PO5/6, PO7/8, Oz, O1/2, and CB1/2; All : 62 channels.) \textbf{b}, The effect of train block size (n =20, mean, error bar is 95\% CI). \textbf{c}, The effect of sub bands (\textbf{\textit{n}}=20, mean, error bar is 95\% CI)}
    \label{fig:enter-label}
\end{figure}

\newpage

\begin{figure}[h]
    \centering
    \includegraphics[scale=1]{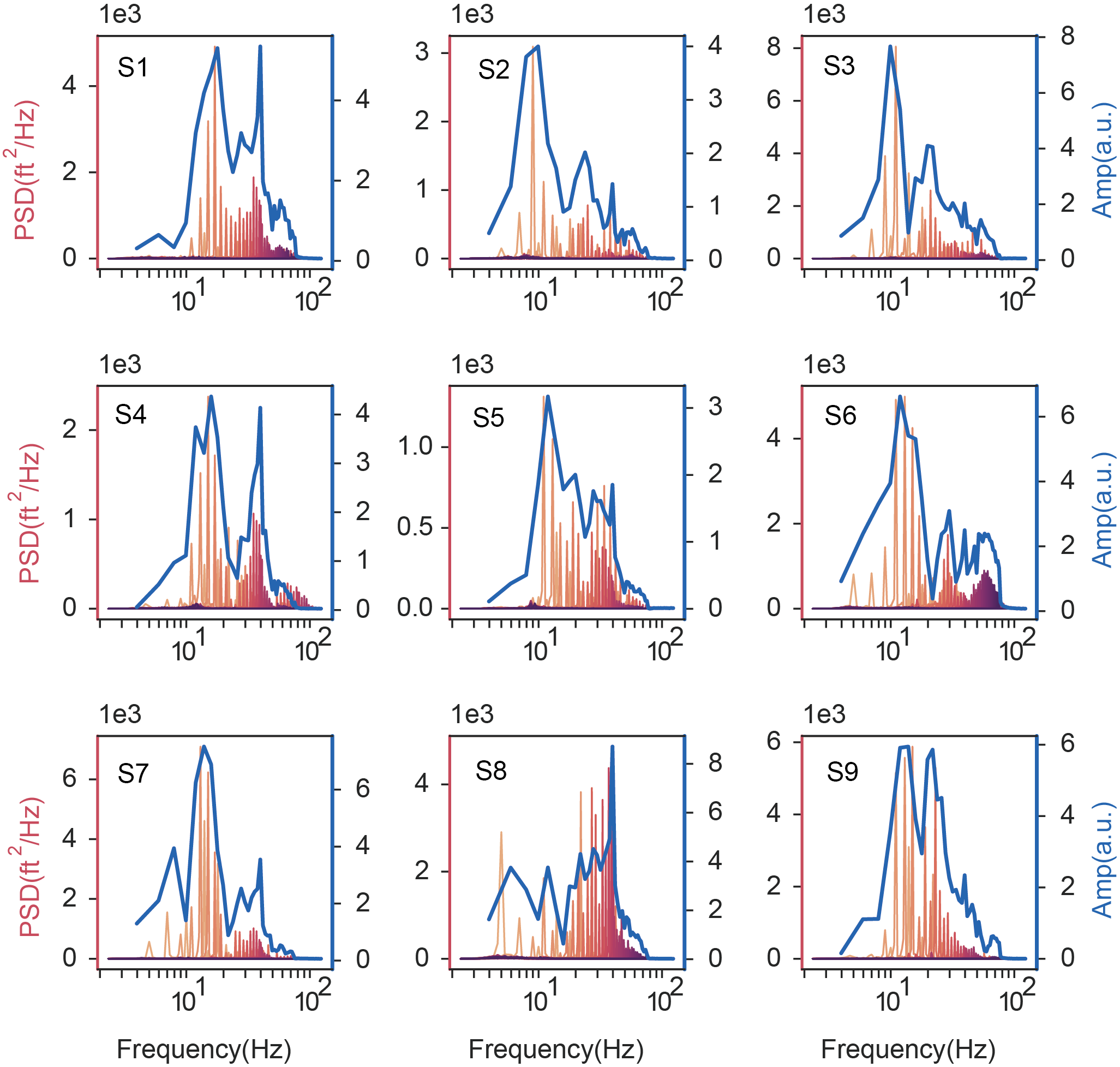}
    \caption{The spectral representation of MEG sweep data for each 9 subjects, 5-75 Hz. Left y-axis represent the PSD of each stimulation frequencies. Right y-axis (blue lines) represents the system’s frequency response $H(f)$.}
    \label{fig:enter-label}
\end{figure}

\newpage

\begin{figure}[h]
    \centering
    \includegraphics{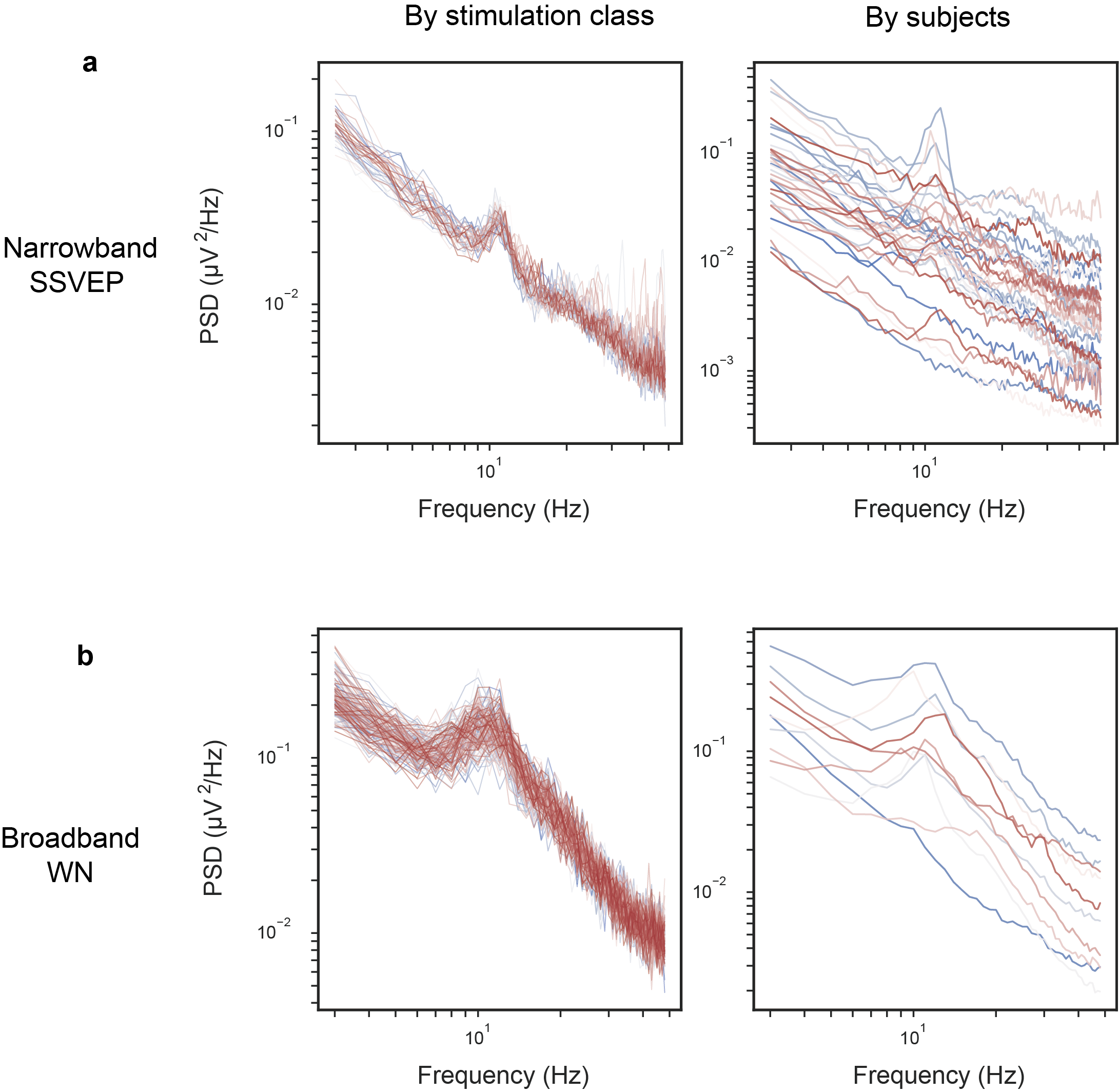}
    \caption{The noise component under both narrowband and broadband stimulation. a, From Wang et al. 2017\cite{wang2016benchmark}, left panel: each line represents a stimulation class (\textbf{\textit{n}}=40), right panel: each line represents a subject (\textbf{\textit{n}}=35) b, Under broadband WN stimulation from preliminary experiment, left panel (\textbf{\textit{n}}=160),right panel (\textbf{\textit{n}}=10)}
    \label{fig:enter-label}
\end{figure}

\newpage

\begin{figure}[h]
    \centering
    \includegraphics{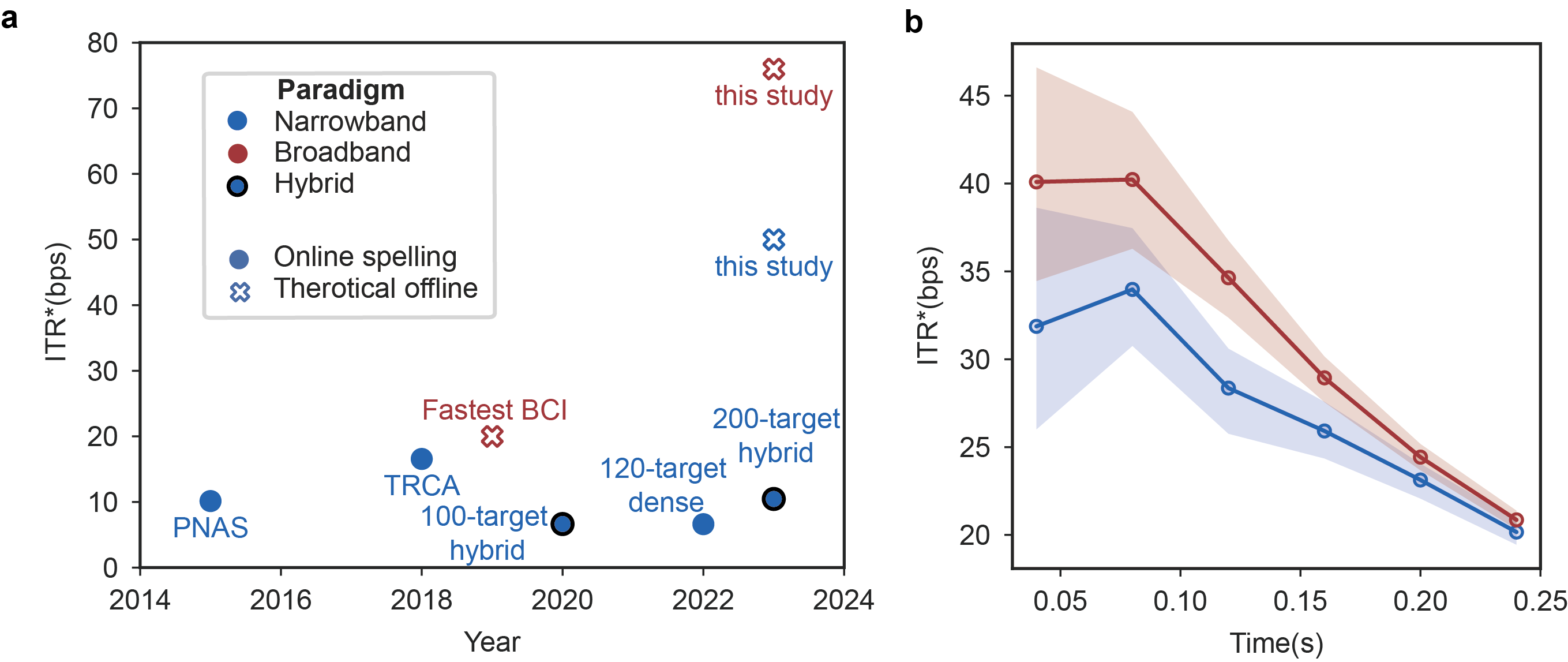}
    \caption{The evolution of ITR* over time. \textbf{a}, The comparison of related studies with the proposed studies. The filled marker represents the performance reported by online spelling. The unfilled ones represent the theoretical value validated by offline stimulation. The related works are: PNAS\cite{chen2015high}, TRCA\cite{nakanishi2017enhancing}, Fastest BCI\cite{nagel2019world}, 100-target hybrid\cite{xu2020implementing}, 120-target dense\cite{chen2022spectrally}, 200-target hybrid\cite{han2023high}, this study (SSVEP and WN) \textbf{b}, The ITR* of two paradigms under time window $<$0.3 s. Results showed that the distinction of two paradigms continue to increase under extreme short time window.}
    \label{fig:enter-label}
\end{figure}

\newpage

\begin{table}[]
\centering
\caption{Mutual information of different visual systems}
\label{tab:my-table}
\begin{tabular}{ccccc}
\hline
Animal & System & Stimulus & I (S,R) (bps) & Method \\ \hline
\multirow{2}{*}{Salamander\cite{warland1997decoding}} & \multirow{2}{*}{RGC} & \multirow{2}{*}{White Noise} & 13.9 & Upper \\
 &  &  & 6.3 & Lower \\
\multirow{2}{*}{Cat\cite{passaglia2004information}} & \multirow{2}{*}{RGC} & \multirow{2}{*}{M-sequence} & 60-100 & Upper \\
 &  &  & 40-60 & Lower \\
\multirow{2}{*}{Guinea pig\cite{freed2005quantal}} & \multirow{2}{*}{RGC} & \multirow{2}{*}{White Noise} & 53+-22 & Upper \\
 &  &  & 13+-3 & Lower \\
\multirow{2}{*}{Macaque\cite{yu2010estimating}} & \multirow{2}{*}{LGN} & \multirow{2}{*}{White Noise} & \multirow{2}{*}{50-120} & \multirow{2}{*}{Upper} \\
 &  &  &  &  \\
\multirow{2}{*}{Rhesus monkeys\cite{buracas1998efficient}} & \multirow{2}{*}{MT} & \multirow{2}{*}{Moving Grating} & 5.5 & Lower \\
 &  &  & 12 & Direct \\
\multirow{2}{*}{Human(this study)} & \multirow{2}{*}{RGC-V1/2} & \multirow{2}{*}{White Noise} & 63+-20 & Upper \\
 &  &  & 25+-3 & Lower \\ \hline
\end{tabular}%
\end{table}

\newpage

\bibliography{sn-bibliography}% common bib file

\end{document}